# Ultrahigh-$Q$ All-Metallic Metasurfaces with Robust Near-Perfect Absorption

Hanlei Xu[1], Han Qi[1], Chaoying Shi[1], Hong Zhang[1], Jun Wang[1], Ziming Meng[2], Guoliang Deng[1*] and Hao Zhou[1*]

[1]*College of Electronics and Information Engineering, Sichuan University, Chengdu 610065, China*

[2]*School of Physics and Optoelectronic Engineering, Guangdong University of Technology, Guangzhou 510006, China*

**Corresponding author(s). E-mail(s): Hao Zhou (zhoufirst@scu.edu.cn);*

*Guoliang Deng (gdeng@scu.edu.cn)*

## Abstract

High quality-factor ($Q$) resonant metasurfaces have attracted significant attention due to their potential applications in cutting-edge fields of optics. However, limited by intrinsic dissipation losses, achieving both an extremely high $Q$ factor and perfect absorption for strong light-matter interaction control in plasmonic metasurface is still highly challenging. Here, we demonstrate a plasmonic metasurface composed of symmetric double-pillars (SDPs) on a gold film, in which the flexibly tuned geometric space enables precise control over the nonlocality of dark-mode Fabry-Pérot bound states in the continuum (FP-BIC) and its coupling with Rayleigh anomaly (RA)-associated lattice resonance. Based on the coupling control between these two modes, the radiative and dissipative losses are well balanced, resulting in a measured $Q$ factor of 2180 (theoretical 2800) and absorption nearly 99%. Notably, the near-perfect absorption response is maintained over broad geometric parameter windows of SDPs (height $h = 70\text{-}120$ nm, radius $r = 210\text{-}280$ nm), while the resonance retains a consistently high $Q$ factor throughout these ranges. This hybrid coupling strategy establishes a general framework for designing high-$Q$, near-critical coupling plasmonic devices for ultrasensitive sensing, narrowband filtering, and active photonics.

## Introduction

Subwavelength metal nanostructures support localized surface plasmon resonances (LSPRs), arising from the coupling of electromagnetic fields with free-electron plasma at a metal-dielectric interface[1, 2]. This enables plasmonic metasurfaces composed of metallic building blocks to confine light and strongly enhance electromagnetic fields, thereby boosting light-matter interactions to unprecedented levels[3, 4, 5, 6]. By leveraging these extraordinary characteristics, plasmonic metasurfaces have driven advances in numerous fields including sensing[7, 8, 9], spectroscopy[10, 11], nanolaser[12] and nonlinear optics[13, 14]. However, owing to the strong radiative damping and intrinsic Ohmic losses, the LSPRs generally suffer from low $Q$ factors[15, 16, 17, 18]. Such a limitation poses a major challenge for plasmonic metasurfaces in scenarios requiring long photon lifetimes and ultranarrow linewidths[19].

To address the low-$Q$ characteristics of plasmonic resonances, recent studies have increasingly turned to the physics of bound states in the continuum (BICs), which offers an effective route to suppress radiative leakage in plasmonic systems[20, 21, 22, 23]. BICs are exotic localized eigenstates embedded in the radiation continuum[24, 25], in ideal situations, destructive interference among radiation channels or symmetry mismatch with external plane waves completely suppresses far-field coupling, thereby giving rise to a dark state with, in principle, an infinite radiative lifetime and a diverging radiative quality factor $Q_{\mathrm{rad}}$. In practice, structural asymmetry, finite-size effects, and fabrication imperfections perturb ideal BICs into quasi-BIC (qBIC) modes with finite yet still remarkably high $Q$ factors[26]. With the extension of the BIC concept from dielectric systems to hybrid metal-dielectric and purely plasmonic platforms, BIC-based design has emerged as an important route for suppressing radiative leakage and accessing substantially enhanced $Q$ factors[27, 28, 29]. Beyond enhancing the radiative quality factor $Q_{\mathrm{rad}}$ through BIC-induced suppression of radiative leakage[25], another important route toward high-$Q$ plasmonic resonances is to reduce dissipative loss through a local-to-nonlocal modal evolution[30, 31, 32]. In this process, the resonant field becomes less confined and more collectively distributed, which alleviates hotspot-induced metal loss and thereby improves the dissipative quality factor $Q_{\text{dis}}$[29].

Relatedly, periodic nanoparticle arrays can support collective high-$Q$ resonances known as surface lattice resonances (SLRs), which originate from the coupling between LSPRs and Rayleigh anomalies (RAs)[33, 34, 35]. Benefiting from collective resonances, SLRs provide a promising route to ultrasharp plasmonic resonances by preserving strong near-field enhancement while suppressing radiative damping[36, 37]. Previous studies have shown that $Q$ factors on the order of $10^3$ can be achieved by engineering the unit-cell geometry and lattice period[38, 39], and experimentally demonstrated SLR with record-high $Q$ factor of 2340 have further highlighted the potential of this approach[36]. Achieving such high-Q SLRs often requires refractive-index matching between the substrate and superstrate[34, 36]. Recent work further demonstrated that high-Q SLRs can also be realized in index-discontinuous environments, with waveguide-assisted dielectric-environment engineering[40]. Moreover, mode hybridization between SLRs and BICs has recently emerged as a possible route to further tailor resonance linewidths and modal properties[41, 42]. Nevertheless, studies on SLR-BIC hybridization remain scarce, and the few reported demonstrations still mainly rely on refractive-index tuning, similar to previous high-$Q$ SLR schemes[41, 43, 44]. The reliance on dielectric-environment engineering imposes additional constraints on device design and limits the flexibility of resonance tuning. Taken together, although BIC-based, SLR-based, and SLR-BIC hybridized approaches can improve $Q$ factors to some extent, the attainable performance remains limited. More importantly, a systematic strategy for simultaneously controlling radiative and dissipative losses is still lacking, making it difficult to achieve $Q_{\text{rad}} \approx Q_{\text{dis}}$ for critical coupling[45]. Therefore, the simultaneous realization of ultrahigh-$Q$ resonances and near-critical coupling in plasmonic metasurfaces remains highly challenging.

In this article, we propose and experimentally demonstrate an all-metal plasmonic metasurface consisting of symmetric double-pillars (SDPs) on a golden film substrate. By employing Rayleigh anomalies (RAs) as radiative channels to couple with dark mode Fabry-Pérot bound states in the continuum (FP-BICs), we realize a high-$Q$ hybrid mode in which radiative and dissipative losses are precisely balanced, enabling near-critical coupling ($Q_{dis} \approx Q_{rad}$) in a lossy metallic system. The metasurface exhibits a simulated $Q$ factor of 2800, an experimental $Q$ factor of 2180, and nearly 99% absorption. To the best of our knowledge, this represents the highest experimentally reported $Q$ factor among all-metallic metasurfaces. Importantly, such a near-critical coupling state is sustained over broad geometric ranges of $h = 70$-$120$ nm and $r = 210$-$280$ nm, throughout which the hybrid resonance remains at a consistently high $Q$ level. These findings establish an experimentally accessible platform for simultaneously achieving ultrahigh-$Q$ resonances and near-critical coupling in all-metallic metasurfaces.

## Results

### Metasurface design and intrinsic modal evolution

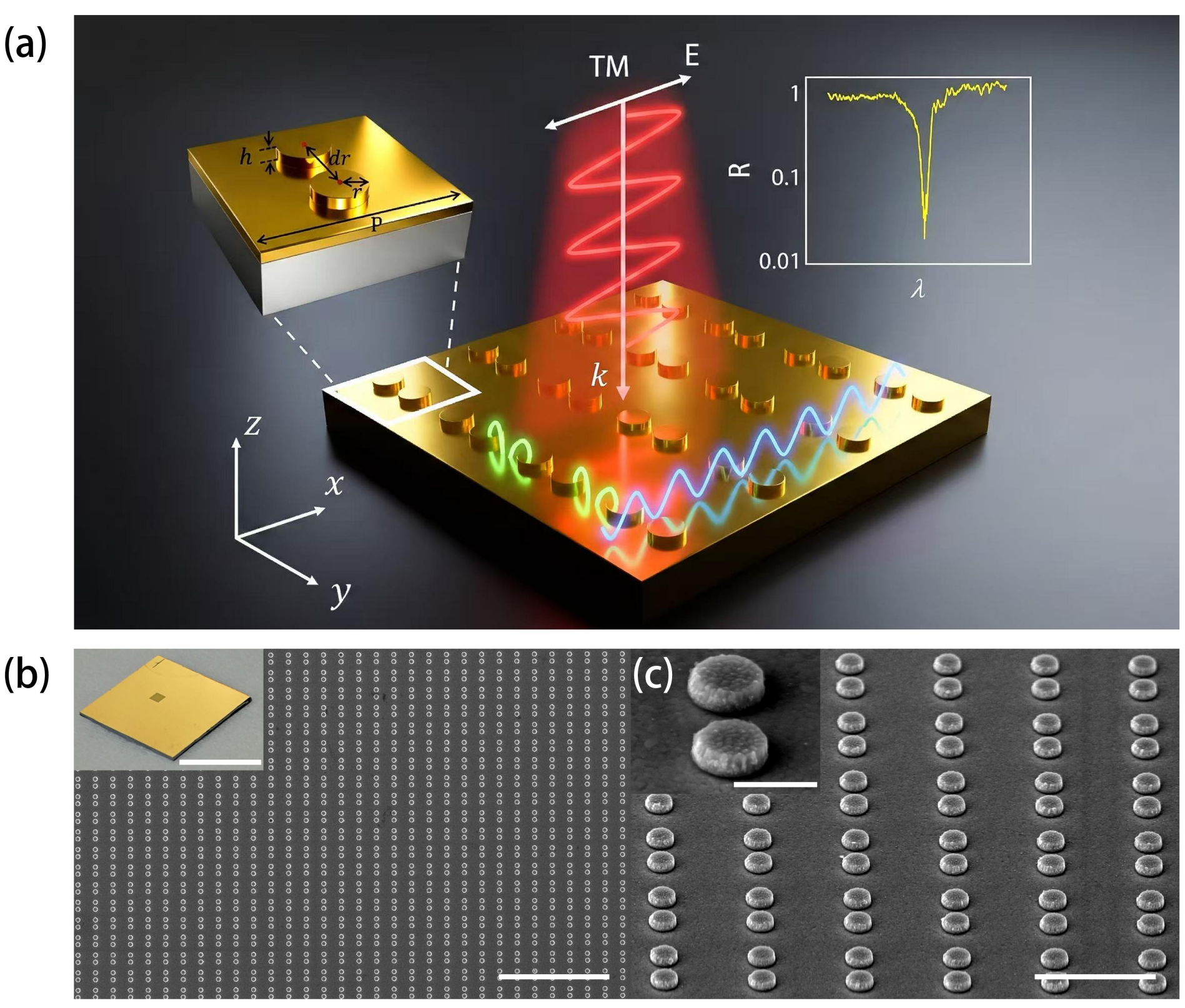

**Figure 1. The symmetric double-pillars array metasurface. (a)** Schematic of the symmetric double-pillars array metasurface. The inset in the upper-left corner shows the geometric parameters of the unit cell, where the pillar height is h=80 nm, radius r=213 nm, center-to-center distance $d_r$=640 nm, and period P=1600 nm, as shown in the figure. The inset in the upper-right corner demonstrates the experimentally achieved ultra-high $Q$ value of 2180 and nearly 99% resonance intensity of the plasmonic metasurface. The blue and green curves represent the RA-associated lattice resonance and the FP-BIC like resonance, respectively. **(b)** Scanning electron microscope (SEM) image of the studied metasurface, scale bar, 10 μm; the inset is sample image, scale bar, 1 cm. **(c)** The right image is a 45° tilted SEM image, scale bar, 2 μm; the inset, scale bar, 500 nm.

The SDPs metasurface studied in this work is schematically shown in Fig. 1(a). The structure is formed by a periodic array of SDPs with same structural parameters on a golden film substrate. The geometric parameters of the unit cell are shown in the upper left corner inset Fig. 1(a). The SDPs with a height of $h$ = 80 nm and a radius of $r$ = 213 nm, are arranged in a square lattice with a periodicity of $P$ = 1600 nm. The center-to-center distance $dr$ between the two Au nanopillars is 640 nm. In this work, the thickness of gold film is 80 nm (typically for thicknesses > 30 nm[28] ) for the metasurface to ensure transmittance T to be 0 in the near-infrared wavelength range due to the optical opacity. The relationship between absorption $A$ and reflection $R$ can be simplified as $R + A = 1$. Figs. 1(b) and 1(c) show the SEM images of the fabricated all-metal SDP metasurface, with an array size of 1000 × 1000 $\mu m^2$. Details of the fabrication process are provided in Methods and Supporting Information S1.

To clarify the modal coupling in the proposed metasurface, we first identify and analyze the two fundamental modes supported by the SDPs metasurface through full-wave finite element method (FEM) simulations. The SDPs metasurface supports bright (odd-symmetry in-plane resonance) and dark (even-symmetry out-of-plane resonance) LSPRs due to plasmonic material system and structural design. In this work, the pillar height $h$ serves as a critical parameter that facilitates a transition between local and nonlocal regimes for both modes. At the Γ point $(k_x=0)$, the bright mode can be directly excited under TE-polarized incidence, whereas the dark mode is symmetry-protected and corresponds to a bound state in the continuum (BIC).

Fig. 2 summarizes how the bright and dark modes evolve from a local regime to a nonlocal regime as $h$ decreases.

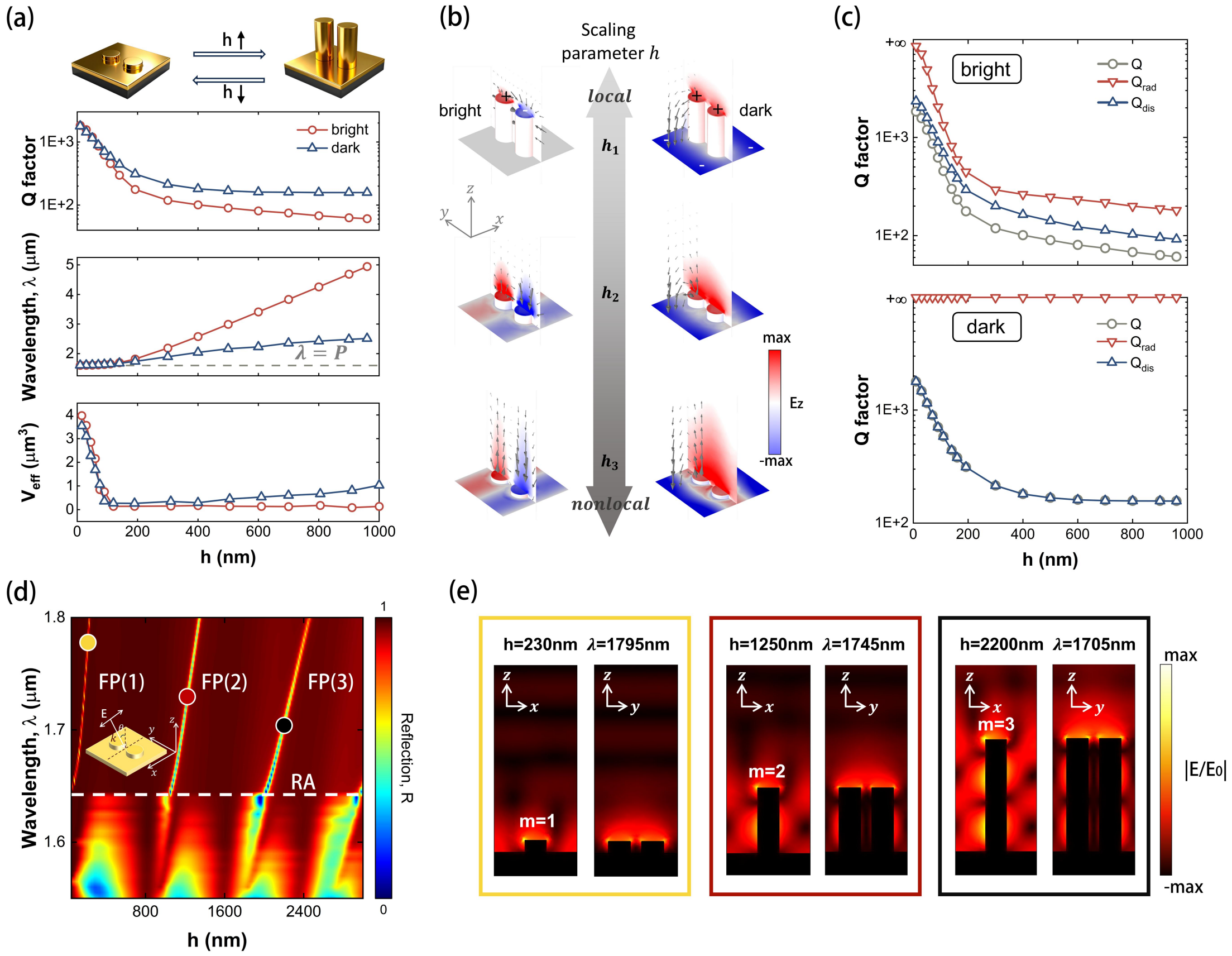

**Figure 2. Local-to-nonlocal modal evolution and Fabry–Pérot-like dark-mode characteristics in the SDPs metasurface. (a)** Transition between local and nonlocal resonances through parameter h scaling. Dependence of $Q$ factor, the effective mode volume $V_{\text{eff}}$ and resonance wavelength $\lambda$ on $h$ for the bright mode and dark mode. **(b)** The corresponding electric field distribution ($E_z$) for various $h$ values are shown, ($h_1 = 960\ nm,\ h_2 = 192\ nm,\ h_3 = 80\ nm$) **(c)** Evolution of the total quality factor $Q$, radiation quality factor $Q_{\text{rad}}$, and dissipation quality factor $Q_{\text{dis}}$ for the dark mode at the Γ point as a function of pillar height h. **(d)** Evolution of reflection spectra for different pillar heights h under 1.5° oblique incidence. Three discrete reflection dips corresponding to Fabry–Pérot-like resonances FP (1), FP (2), and FP (3) are shown, with the white dashed line representing the (-1,0) order RA. **(e)** Electric field distributions for different orders of FP1, FP2, and FP3. The three differently colored boxes correspond to the resonance peaks in Figure (c) at the positions of the circles.

Fig. 2(a) presents the dependence of the total quality factor $Q$, resonant wavelength $\lambda$, and effective mode volume $V_{\text{eff}}$ of the bright and dark modes on the pillar height $h$ at $k_x = 0$. Here, the effective mode volume is defined as $V_{\text{eff}} =$ $\int W(\mathbf{r})\, dV/W_{\max}$, where $W(\mathbf{r})$ is the electromagnetic energy density and $W_{\max}$ is its maximum value[46]. Physically, a smaller $V_{\text{eff}}$ indicates stronger field confinement near metallic hotspots, leading to a larger field overlap with the lossy metal and hence stronger Ohmic dissipation, and thus a lower $Q_{\text{dis}}$. Correspondingly, the dissipative quality factor satisfies

$$Q_{\text{dis}} \propto \left(\int_{V_{\text{metal}}} \text{Im}\,(\varepsilon_m)|E|^2\, dV\right)^{-1} \tag{1}$$

as $h$ decreases, $V_{\text{eff}}$ of both modes increases markedly, while their resonant wavelengths gradually approach the lattice period $P$. This spectral evolution reflects the transition from a geometry-dominated local regime to a lattice-influenced nonlocal regime: in the local regime, the resonance is mainly determined by the individual pillar geometry, whereas in the nonlocal regime, the mode becomes increasingly governed by inter-unit coupling and lattice feedback, causing its wavelength to move closer to the periodicity-associated spectral scale. Meanwhile, the total $Q$ factors of both modes are substantially enhanced, indicating suppressed dissipative loss and strengthened nonlocality.

The physical origin of this transition is directly visualized in Fig. 2(b), which shows the $E_z$ field distributions of the bright and dark modes for three representative heights ($h_1 = 960$ nm, $h_2 = 192$ nm, $h_3 = 80$ nm). For large $h$, both modes exhibit strongly localized fields concentrated near the pillar tops, characteristic of a local regime where the response is mainly confined within individual unit cells. As $h$ decreases, the hotspot confinement is gradually weakened, and the fields extend into the air regions between neighboring unit cells and over the golden film plane, indicating a nonlocal regime with enhanced in-plane coupling and coherent field sharing among multiple unit cells.

Fig. 2(c) further quantifies how this local-to-nonlocal evolution affects the loss channels. The total $Q$ factor consists of the radiative quality factor $Q_{\text{rad}}$ and dissipative quality factor $Q_{\text{dis}}$, i.e., $Q^{-1} = Q_{rad}^{-1} + Q_{dis}^{-1}$, with $Q = \omega_r/2\omega_i$ determined from the complex eigenfrequency $\omega = \omega_r + i\omega_i$, where $\omega_r$ and $\omega_i$ are the real and imaginary parts of eigenfrequencies. Consistent with Eq. (1), $Q_{\text{dis}}$ for both modes increases substantially with decreasing $h$, owing to the enlarged $V_{\text{eff}}$ and reduced field overlap with the metal. For the bright mode, $Q_{\text{rad}}$ also increases but remains finite, so the total $Q$ is still limited mainly by dissipation. For the dark mode, radiative leakage is symmetry-forbidden at the $\Gamma$ point, yielding an ideal BIC with $Q_{\text{rad}} \to \infty$. As a result, the enhancement of the dark-mode $Q$ is dominated by the increase in $Q_{\text{dis}}$. As this work mainly focuses on the dark mode, whose radiative loss is strongly suppressed, a detailed discussion of the bright mode is provided in Supporting Information S3.

To further clarify the physical origin of the dark-mode BIC, we examine the reflection spectra under TM-polarized oblique incidence, where the in-plane wave vector is $k_x = \frac{2\pi}{\lambda}\sin(1.5°)$=$0.1645/\lambda$. As shown in Fig. 2(d), several narrow reflection-dip branches emerge and evolve with the pillar height $h$. The white dashed line marks the Rayleigh anomaly (RA) boundary of the $(-1,0)$ diffraction order, whose wavelength is given by

$$\lambda_{\text{RA}}^{(m,\,n)}(k_x) = \frac{2\pi n_{air}}{\sqrt{(k_x+\frac{2\pi m}{P})^2+(\frac{2\pi n}{P})^2}} \tag{2}$$

where $P$ is the lattice period, $(m, n)$ is the diffraction order and $n_{air}$ is the refractive index of air. With increasing $h$, three reflection-dip branches exhibit pronounced redshifts, indicating strong structural-height dependence. It is evident that three bands of reflectance dip shift with increasing the height $h$. These bands correspond to Fabry-Pérot -like (FP-like) modes of different orders, which can be described by the FP resonance condition[44]: $\beta \cdot 2H + \phi_r = m \cdot 2\pi$, where $\beta$ is the propagation constant of the surface plasmon polariton (SPP), $H$ is the effective cavity height, $\phi_r$ is the round-trip reflection phase, and $m$ is the FP order.

Fig. 2(e) shows the electric field distributions at the marked resonant frequencies under 1.5° TM-polarized oblique incidence, corresponding to the yellow, red, and black circle markers in Fig. 2(d), i.e., $(h, \lambda) = (230\text{ nm}, 1795\text{ nm})$, $(1250nm, 1745nm)$, and $(2200nm, 1705nm)$, respectively. The three modes exhibit clear standing-wave patterns along the vertical cavity, corresponding to the first, second, and third order FP-like resonances. Specifically, surface plasmon polaritons (SPPs) propagate vertically and are reflected between the SDP tops and the gold substrate, thereby forming an open and dissipative FP cavity in the coupled-pillar region. These results confirm that the dark mode originates from an FP-BIC mechanism.

Taken together, Fig. 2 shows that decreasing $h$ drives both bright mode and dark mode toward a more nonlocal regime and significantly enhances $Q_{\text{dis}}$. However, critical coupling cannot be achieved through this intrinsic modal evolution alone. For the bright mode, $Q_{\text{rad}}$ and $Q_{\text{dis}}$ vary in a similar manner with $h$, preventing their balance. For the dark mode, its BIC nature at the $\Gamma$ point enforces $Q_{\text{rad}} \to \infty$. Therefore, although the local-to-nonlocal evolution is highly effective in increasing the overall $Q$ factor, it still does not provide access to the critical-coupling condition $Q_{\text{rad}} \approx Q_{\text{dis}}$. In this sense, the pillar height $h$ serves as a key parameter for controlling modal nonlocality, dissipative loss, and the radiative leakage of the FP-BIC.

## Hybridization between the FP-BIC and RA modes

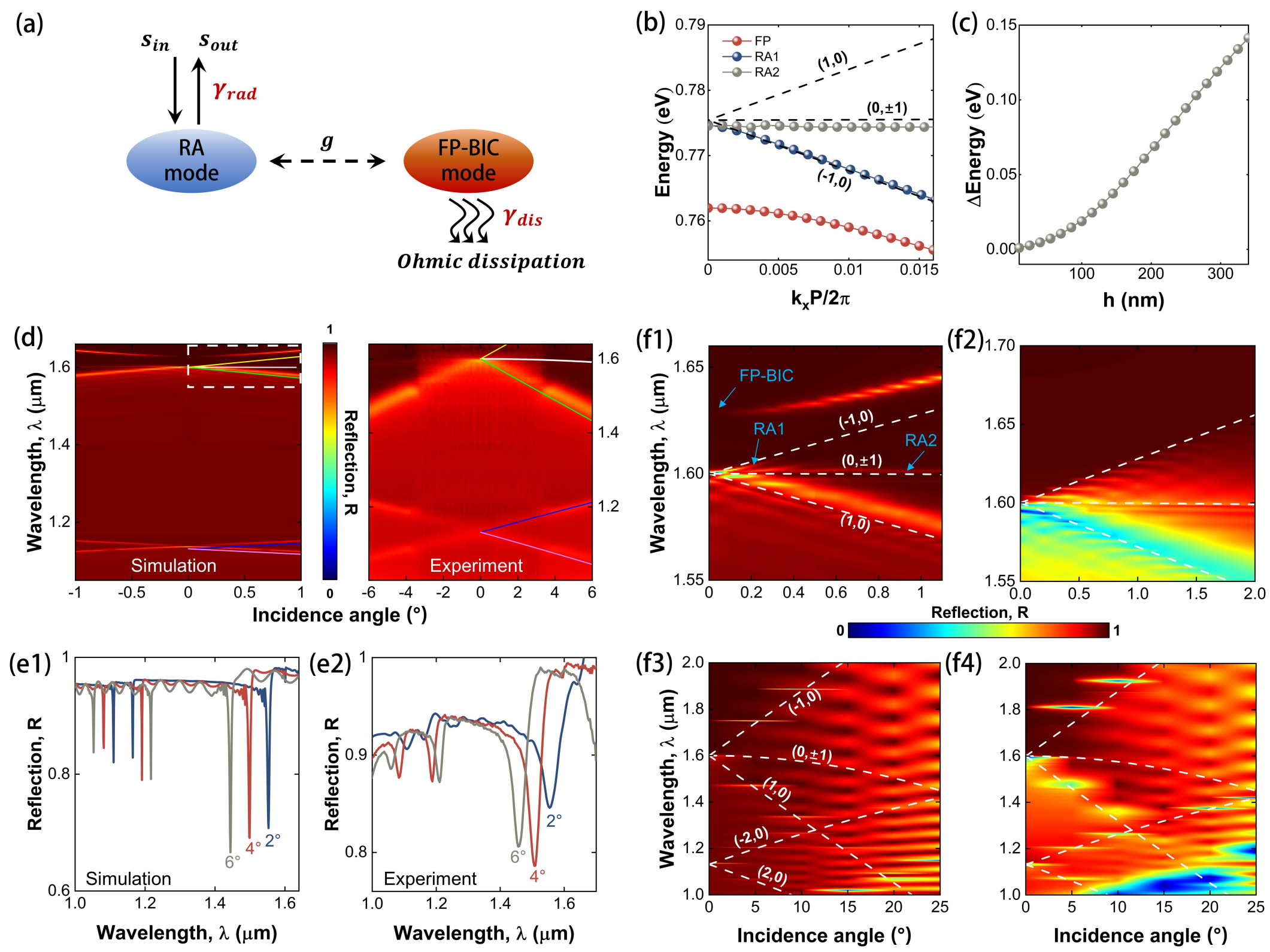

**Figure 3. Physical image and dispersion characteristics of nonlocal multi-mode coupling. (a)** Schematic of two-mode coupling based on the Time-domain Coupled-mode Theory (TCMT). The lattice radiation resonance related to the Rayleigh anomaly (RA) couples to the single reflection port ($s_{\mathrm{in}}, s_{\mathrm{out}}$) via the external coupling rate $\gamma_{\mathrm{rad}}$; the FP-BIC dark mode is primarily limited by metal ohmic dissipation $\gamma_{\mathrm{dis}}$. **(b)** Calculated band structure near the Γ point for h=80nm, with the black dashed lines representing the diffraction boundaries. **(c)** The frequency detuning $\Delta E = |E_{\text{FP-BIC}} - E_{\mathrm{RA}}|$ between FP-BIC and RA modes at the Γ point as a function of $h$. **(d)** Angle-resolved reflection spectra under TM polarization at h=80nm, with simulation on the left and experiment on the right. The colored curves represent different Rayleigh anomaly orders: (-1, 0) in yellow, (0, ±1) in white, (1, 0) in green, (-2, 0) in blue, and (2, 0) in pink. **(e1)** Simulation and **(e2)** experiment of reflection spectra under different incident angles, with the incident angles indicated in the figure. **(f)** Angle-resolved reflection spectra under TM polarization for different heights and angle ranges: **(f1)** and **(f3)** for $h$=80, **(f2)** and **(f4)** for $h$=200nm. The white dashed lines represent diffraction orders, and **(f1)** is a magnified view of the white dashed box in **(d)**.

Although the FP-BIC supports a high $Q$ factor, its strongly suppressed radiative leakage hinders efficient energy coupling and thus limits practical functionality. To overcome this limitation, we intentionally couple the FP-BIC mode to a RA-associated lattice resonance, as schematically illustrated in Fig. 3(a). The introduction of RA mode provides an external radiative channel. Their hybridization opens a controllable radiation pathway for the dark mode while retaining its high-$Q$ character. In this framework, the RA mode mainly provides an external radiative channel, whereas the FP-BIC dark mode is dominated by Ohmic dissipation, and their coherent interaction is characterized by a coupling strength $g$. By tailoring this hybridization, the balance between radiative and dissipative losses can be continuously tuned, enabling near-critical coupling and tunable perfect absorption. Near the Γ point, the coupled system can be described by a simplified non-Hermitian Hamiltonian[47]:

$$H = \begin{bmatrix} \omega_{RA-i\gamma_{RA}} & g \\ g & \omega_{FP-i\gamma_{FP}} \end{bmatrix} \tag{3}$$

here, $\omega_{\mathrm{RA}}$ and $\omega_{\mathrm{FP}}$ denote the eigenfrequencies of the RA-associated lattice resonance and FP-BIC mode, respectively, while $\gamma_{\mathrm{RA}}$ and $\gamma_{\mathrm{FP}}$ are their corresponding decay rates. The resulting isolated hybrid resonance can be further described by a one-port temporal coupled-mode theory (TCMT) model[48]:

$$\frac{da}{dt} = (i\omega_0 - \gamma_{\mathrm{rad}} - \gamma_{\mathrm{dis}})a + \sqrt{2\gamma_{\mathrm{rad}}}\, s_{\mathrm{in}} \tag{4}$$

$$s_{\mathrm{out}} = s_{\mathrm{in}} - \sqrt{2\gamma_{\mathrm{rad}}}\, a \tag{5}$$

where $a$ is the modal amplitude of the hybrid resonance, $s_{\mathrm{in}}$ and $s_{\mathrm{out}}$ are the amplitudes of the incident and reflected waves, respectively, $\omega_0$ is the resonant angular frequency, and $\gamma_{\mathrm{rad}}$ and $\gamma_{\mathrm{dis}}$ denote the radiative and dissipative decay rates, respectively. Since the bottom gold film fully suppresses transmission, the metasurface operates as a one-port resonator, with reflection coefficient

$$r(\omega) = \frac{s_{\mathrm{out}}}{s_{\mathrm{in}}} = 1 - \frac{2\gamma_{\mathrm{rad}}}{i(\omega-\omega_0)+\gamma_{\mathrm{rad}}+\gamma_{\mathrm{dis}}} \tag{6}$$

at $\omega = \omega_0$, the reflection vanishes when $\gamma_{\mathrm{rad}} = \gamma_{\mathrm{dis}}$, corresponding to the critical-coupling condition. Therefore, the essential role of mode hybridization is to tune the radiative loss toward the dissipative loss, thereby enabling a high-$Q$ hybrid resonance in the near-critical coupling regime.

We first analyze this coupling from the calculated band structure. Fig. 3(b) shows the calculated band structure of the metasurface at $h = 80$ nm, where the FP-BIC branch lies close to two RA-associated branches, denoted as RA1 and RA2, near the Γ point around $\lambda \approx 1.6\, \mu m\, (0.775\, eV)$. The dashed lines mark the corresponding diffraction boundaries, indicating that the interaction occurs in the vicinity of the RA channels. Near the Γ point, the FP-BIC and RA branches are spectrally close, providing favorable conditions for their hybridization. Fig. 3(c) further reveals how this coupling can be controlled by the pillar height $h$. To characterize the coupling condition, we plot the energy detuning

$$\Delta E = |E_{\mathrm{FP}} - E_{\mathrm{RA}}| \tag{7}$$

as a function of the $h$. As $h$ increases, $\Delta E$ increases monotonically, whereas reducing $h$ continuously brings the FP-BIC and RA modes closer in energy. This result indicates that the pillar height provides an effective handle for tuning the spectral overlap and thus the hybridization condition between the two modes. Physically, the RA branch is mainly governed by the lattice condition and is therefore only weakly dependent on $h$, whereas the FP-BIC branch is highly sensitive to the effective vertical cavity height and can be shifted efficiently by varying $h$.

We confirmed the band analysis through angle-resolved reflection spectroscopy at $h = 80$ nm. Fig. 3(d) shows the simulated (left) and experimental (right) reflection dispersions of the SDPs metasurface under TM-polarized incidence. The colored curves denote the RA boundaries of different diffraction orders, including (-1, 0) in yellow, (0, ±1) in white, (1, 0) in green, (-2, 0) in blue, and (2, 0) in pink. Several resonant branches closely follow these RA boundaries over a broad angular range, indicating that the observed modes are strongly influenced by the RA-associated diffraction channels. Near normal incidence, the FP-BIC branch approaches the RA-associated lattice resonance branches and hybridizes with them to form an ultrahigh-$Q$ resonance near 1.6 $\mu m$, accompanied by a pronounced reflection dip. A higher-order hybridized feature is also observed near 1.1 $\mu m$. The experimentally measured modal positions agree well with the simulated ones over the accessible spectral range. The corresponding reflection spectra extracted at selected incident angles are shown in Figs. 3(e1, e2), where the good agreement between experiment and simulation further supports the reliability of the numerical model.

The role of pillar height in tuning the hybridization is clarified in Figs. 3(f1)-3(f4). Here, Fig. 3(f1) is a magnified view of the white dashed box in Fig. 3(d), highlighting the small-angle spectral region near the hybridized resonance for $h = 80$nm; Fig. 3(f2) shows the corresponding small-angle dispersion for $h = 200$nm, while Figs. 3(f3) and 3(f4) present the respective wider-angle spectra. For $h = 80$ nm, the FP-BIC lies very close to the RA branches, and the hybridized spectral feature develops in the immediate vicinity of the (-1, 0) and (0, ±1) RA boundaries around 1.6 $\mu m$. In contrast, when $h$ increases to 200 nm, the FP-BIC branch redshifts away from this spectral window, so its direct spectral proximity to the nearby RA branches is no longer visible in the magnified range, and the hybridized feature becomes much less pronounced. This behavior is fully consistent with the increased detuning extracted in Fig. 3(c). Figs .3(f3) and 3(f4) further reveal the global dispersive behavior over a wider angular range for $h = 80$ and $h = 200$ nm, respectively. In both cases, the observed resonant branches evolve closely along the RA boundaries, demonstrating that their overall dispersion is fundamentally governed by the RA-associated diffraction channels. Therefore, the wide-angle spectra establish the RA-related nature of these modes, whereas the small-angle spectra directly show how the $h$-dependent detuning between FP-BIC and RA controls the degree of hybridization near the Γ point. Taken together with the band-structure and detuning analyses, these results confirm that decreasing $h$ progressively improves the hybridization condition between the FP-BIC and RA, which is required for the ultrahigh-$Q$, strongly absorbing hybrid resonance.

## Robust near-critical coupling and high-$Q$ resonances

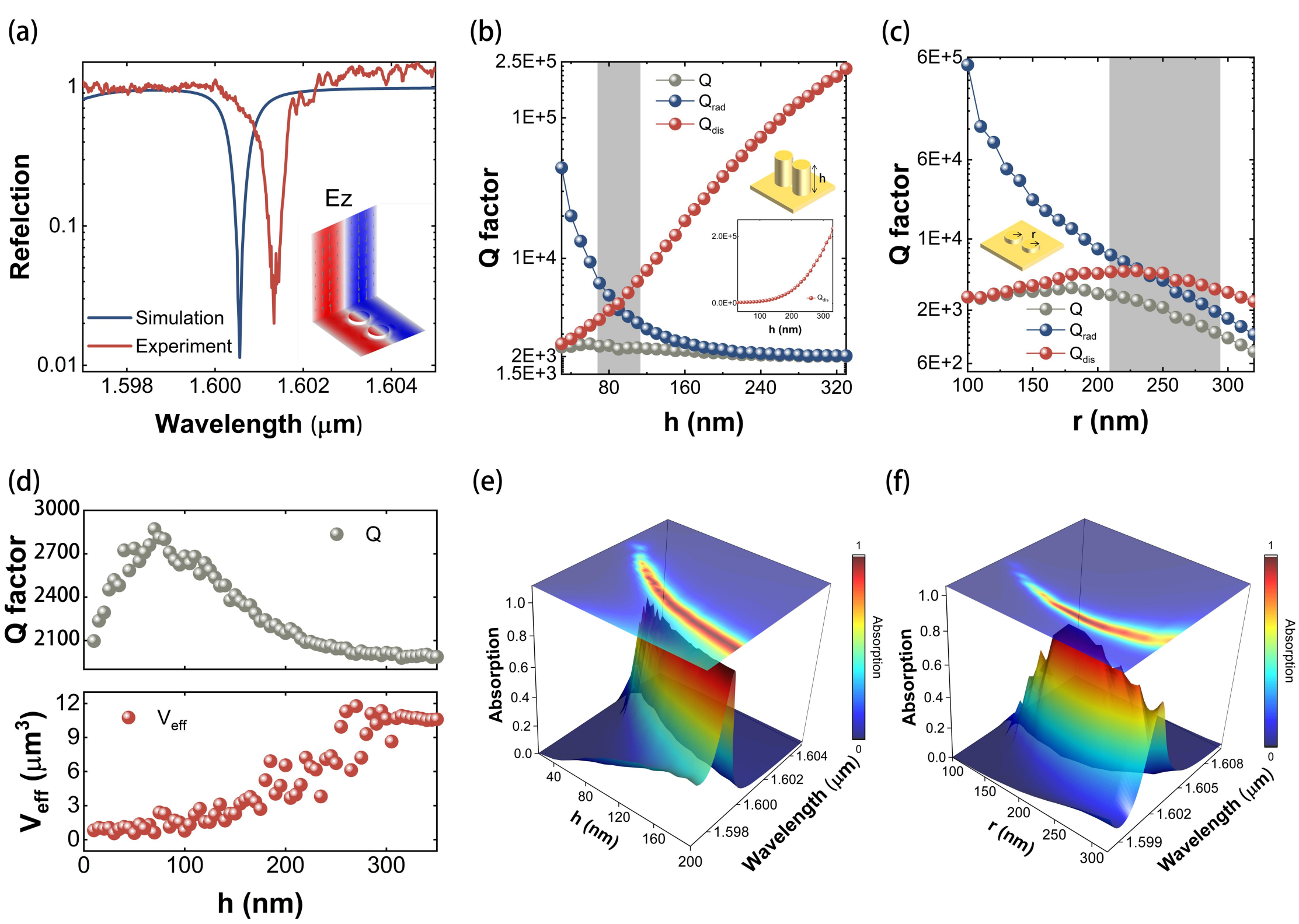

**Figure 4. Near-critical coupling and robust perfect absorption of the ultrahigh-$Q$ hybrid mode.** **(a)** Reflection spectra under TM incidence at h=80nm, with simulation and experimental results. The inset shows the spatial distribution of $E_z$ at the resonance. **(b, c)** Decomposition of the Γ-point quality factor based on complex eigenfrequency extraction: total quality factor $Q$ and its radiation component $Q_{\mathrm{rad}}$, and dissipation component $Q_{\mathrm{dis}}$ as a function of geometric parameters. **(b)** as a function of pillar height h, **(c)** as a function of pillar radius r. The vertical axis is on a logarithmic scale, and the gray area indicates the near-critical coupling region. The inset in **(b)** shows $Q_{\mathrm{dis}}$ as a function of h on a linear scale. **(d)** Evolution of the total quality factor $Q$ and effective mode volume $V_{\mathrm{eff}}$ at the Γ point as a function of pillar height h. **(e)** Absorption spectra of the high-$Q$ hybrid coupled mode at the Γ point as a function of pillar height h. **(f)** Absorption spectra of the high-$Q$ hybrid coupled mode at the Γ point as a function of pillar radius $r$.

By tuning the geometric parameters of the SDPs metasurface, the coupled system can be driven into a special regime in which $Q_{\mathrm{rad}}$ and $Q_{\mathrm{dis}}$ become comparable, thereby enabling an ultrahigh-$Q$ resonance together with strong absorption. Based on this design principle, we experimentally and numerically demonstrate an SDPs metasurface supporting an ultrahigh-$Q$ resonance with near-perfect absorption. The fabricated sample ( $h = 80$ nm) was characterized using a high-resolution spectroscopy system (see Supporting Information S2). Fig. 4(a) shows the simulated and experimental reflection spectra of the SDPs metasurface under TM-polarized normal incidence. The simulation yields a $Q$ factor of 2800, while the measured $Q$ factor is 2180. To the best of our knowledge, this represents the highest experimental $Q$ factor reported to date for all-metal reflective plasmonic metasurfaces. As shown in the inset of Fig. 4(a), the $E_z$ field distribution at resonance reveals pronounced collective behavior and strong nonlocality of the corresponding mode. Meanwhile, the reflection minimum approaches zero at the resonance frequency, corresponding to near-perfect absorption in the present one-port configuration, where the bottom gold film completely suppresses transmission. The simulated spectrum gives a full width at half maximum of $\sim 0.73$nm and a peak absorption close to 99%, in good agreement with the experimentally measured resonance position and lineshape. Using the one-port TCMT framework described by Eq. (6), this high-$Q$ resonance with near-perfect absorption can be understood as a hybrid mode formed through the coupling between the RA-associated lattice resonance and the FP-BIC dark mode near the Γ point, where the radiative and dissipative decay channels are brought into close balance.

Fig. 4(b) and 4(c) show the evolution of the total quality factor $Q$, radiative quality factor $Q_{\mathrm{rad}}$, and dissipative quality factor $Q_{\mathrm{dis}}$ of the hybrid mode at the Γ point as functions of the pillar height $h$ and gold nanopillar radius $r$, respectively. According to the one-port TCMT model, the on-resonance absorbance is given by[49]

$$\mathrm{A} = 1 - \left[\frac{\gamma_{\mathrm{dis}}-\gamma_{\mathrm{rad}}}{\gamma_{\mathrm{dis}}+\gamma_{\mathrm{rad}}}\right]^2 \quad (8)$$

When $Q_{\mathrm{rad}}/Q_{\mathrm{dis}} \in [0.6,1.6]$, the absorption exceeds 95%. In this work, we define this interval as the near critical coupling regime, which is highlighted by the gray-shaded regions in Figs. 4(b) and 4(c). Correspondingly, the near-critical coupling condition is very robust, exhibiting a very large fabrication error tolerance, with $h$ ranging from 70 to 120 nm and $r$ ranging from 210 to 280 nm, within which the total $Q$ factor remains above 2600 across the height $h$ window and above 1400 across the radius $r$ window, although it gradually decreases with increasing $r$. The inset of Fig. 4(b) further plots the evolution of $Q_{\mathrm{dis}}$ with increasing $h$ on a linear scale. The observed variation is consistent with the trends in the detuning $\Delta E$ in Fig. 3(c) and the effective mode volume $V_{\mathrm{eff}}$ in Fig. 4(d), indicating that the enhancement of $Q_{\mathrm{dis}}$ is closely associated with the reduced detuning between the dark mode and the RA-associated resonance, together with the increased modal nonlocality.

Fig. 4(d) further presents the evolution of the hybrid mode $Q$ factor and $V_{eff}$ as a function of $h$. When $h < 60\ nm$, the $Q$ factor is mainly limited by $Q_{\mathrm{dis}}$, resulting in a relatively low value. As $h$ increases, the enlarged modal extent and weakened hotspot confinement enhance $Q_{\mathrm{dis}}$, while the hybrid mode simultaneously approaches the near-critical coupling regime around $h \sim 80$ nm, where $Q_{\mathrm{rad}} \approx Q_{\mathrm{dis}}$. In this regime, the system does not necessarily exhibit the absolute maximum total $Q$, but it enables the simultaneous coexistence of an ultrahigh-$Q$ resonance and strong absorption. With a further increase in $h$, the $Q$ factor becomes gradually dominated by $Q_{\mathrm{rad}}$. This transition from a $Q_{\mathrm{dis}}$-dominated to a $Q_{\mathrm{rad}}$-dominated resonance aligns with the increasing trend of $V_{eff}$ with $h$. Figs. 4(e) and 4(f) show the evolution of the hybrid mode's absorption spectra with $h$ and $r$, respectively. Benefiting from the aforementioned high-$Q$ mechanism and realization of the critical coupling condition, the proposed SDPs metasurface remains in a near-critical coupling regime even under significant variations in geometric parameters ( $h \in [70\,nm, 120\,nm]$, $r \in [210\,nm, 280\,nm]$ ). Within these parameter ranges, the metasurface consistently supports ultrahigh-$Q$ resonances together with near-perfect absorption, indicating a broad geometric tolerance of the near-critical coupling response.

## Discussion

This work demonstrates that RA-assisted hybridization provides an effective route to reconcile ultrahigh-$Q$ resonances with near-critical coupling in lossy all-metal plasmonic metasurfaces. By using the RA-associated lattice resonance as a controllable radiative channel for the dark FP-BIC mode, the radiative and dissipative losses can be balanced while preserving the high-$Q$ character of the resonance. The key physical advance lies in separating the roles of energy storage and radiative access: the FP-BIC mode provides a radiation-suppressed high-$Q$ reservoir, whereas the RA channel introduces tunable radiative leakage.

Compared with conventional plasmonic BIC or SLR approaches, the present strategy does not rely primarily on refractive-index matching or dielectric-environment engineering. Instead, the hybridization is controlled through the structural parameters of an all-metal metasurface, providing a more direct handle for balancing $Q_{\mathrm{rad}}$ and $Q_{\mathrm{dis}}$. This structural control allows the resonance linewidth and absorption depth to be engineered simultaneously, which is difficult to achieve using either intrinsic BIC modes or conventional SLRs alone.

We further benchmark the proposed metasurface against representative plasmonic metasurfaces reported previously, as summarized in Fig. 5. Using the experimental $Q$ factor and modulation depth $\Delta I$ as two key metrics, the proposed structure occupies a favorable position in the comparison map, indicating the simultaneous realization of an ultrahigh-$Q$ resonance and near-unity modulation depth. More detailed comparison parameters are provided in Supporting Information S5. Together with the refractive-index sensing results in aqueous glucose solutions provided in Supporting Information S4, these results highlight the potential of the platform for practical light-matter interaction applications.

The combination of ultrahigh $Q$, strong absorption, and geometric robustness makes the proposed metasurface promising for applications that require enhanced light-matter interactions. Future work will explore its use in low-threshold plasmonic lasing, surface-enhanced Raman spectroscopy, nonlinear optical enhancement, narrowband photodetection, and active photonic modulation. Overall, this work establishes a structurally tunable hybrid-coupling framework for realizing robust ultrahigh-$Q$, near-critical-coupling plasmonic metasurfaces.

Overall, this work establishes a structurally tunable hybrid-coupling framework for realizing ultrahigh-$Q$, near-critical coupling plasmonic metasurfaces. By combining local-to-nonlocal modal evolution with RA-assisted coupling of the dark FP-BIC mode, the proposed all-metal SDP metasurface enables simultaneous suppression of dissipative loss and controlled radiative leakage, leading to robust high-$Q$ absorption. This strategy opens opportunities for ultrasensitive refractive-index sensing, narrowband filtering, surface-enhanced spectroscopy, low-threshold plasmonic lasing, and active photonic modulation.

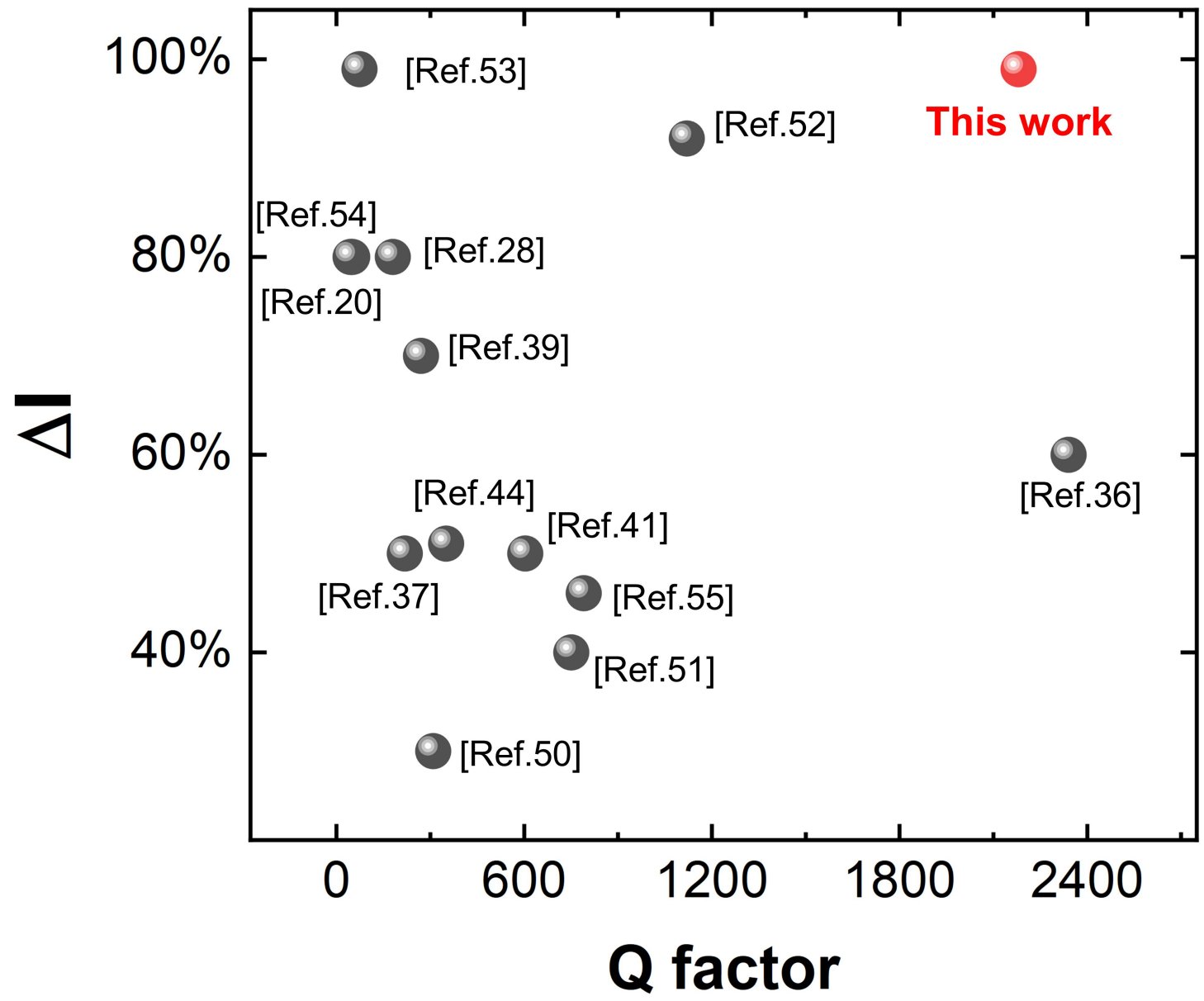

**Figure 5. Comparison of experimental $Q$ factor and modulation depth $\Delta I$ of coupled resonance peaks with existing typical plasmonic metasurface works**[20, 28, 36, 37, 39, 41, 44, 50, 51, 52, 53, 54, 55]**.**

# Methods

## Numerical simulations

Full-wave numerical simulations were performed using the finite element method (FEM). The simulation model consisted of one periodic unit cell of the SDPs metasurface. The lateral boundaries along the $x$- and $y$-directions were set as periodic boundary conditions, while perfectly matched layers (PMLs) terminated by scattering boundary conditions were applied along the $z$-direction to absorb outgoing waves. The metasurface consisted of symmetric double gold pillars placed on a continuous gold film with a thickness of 1000 nm, which was sufficiently thick to suppress transmission. The dielectric function of gold was taken from the built-in material database, using the evaporated gold data from Olmon et al (2012).

For reflection simulations, a plane wave was incident from the air side with the specified polarization and incident angle. The in-plane wave vector was introduced through the Floquet periodic boundary condition. Eigenmode simulations were performed to extract the complex eigenfrequencies and field distributions of the bright, dark, and hybrid modes. The mesh was refined around the gold pillars and the metal-dielectric interfaces to ensure convergence of the resonance wavelength and quality factor.

## Sample fabrication

The SDPs metasurfaces were fabricated on gold-coated silicon substrates using electron-beam lithography, thermal evaporation, and lift-off. First, an 80-nm-thick gold film was deposited on a silicon substrate by thermal evaporation. After electron-beam lithography and development, a second 80-nm-thick gold layer was deposited to form the double-pillar structures, followed by lift-off. Detailed fabrication procedures are provided in Supporting Information S1. The morphology of the fabricated samples was characterized by scanning electron microscopy.

**Optical characterization**

High-resolution reflection spectra near normal incidence were measured using a home-built optical spectroscopy setup, as shown in Supporting Information S2. The reflected signal was recorded by an optical spectrum analyzer and normalized to a reference gold mirror measured under the same conditions. Since the bottom gold film blocks transmission, the absorption was calculated as $A = 1 - R$. Angle-resolved reflection spectra were additionally collected using a commercial angle-resolved spectroscopy system under TE- or TM-polarized incidence to characterize the angular dispersion of the resonances.

**Data processing**

The experimental quality factor was extracted from the resonance linewidth using $Q = \lambda_0/\Delta\lambda$, where $\lambda_0$ is the resonance wavelength and $\Delta\lambda$ is the full width at half maximum obtained from spectral fitting. The modulation depth was defined as $\Delta I = (R_{\mathrm{bg}} - R_{\mathrm{dip}})/R_{\mathrm{bg}}$, where $R_{\mathrm{dip}}$ is the reflectance at resonance and $R_{\mathrm{bg}}$ is the local off-resonant background. The near-critical coupling regime was defined as $Q_{\mathrm{rad}}/Q_{\mathrm{dis}} \in [0.6,1.6]$, corresponding to an on-resonance absorption higher than 95% according to the one-port TCMT model. For refractive-index sensing, the bulk sensitivity was obtained by linear fitting of the resonance wavelength shift as a function of the environmental refractive index, $S = \Delta\lambda_{\mathrm{res}}/\Delta n_{\mathrm{env}}$.

## Data availability

The data that support the findings of this study are available from the corresponding author upon reasonable request. Additional supporting data are provided in the Supplementary Information file.

## Reference

1. Kauranen M, Zayats AV. Nonlinear plasmonics. *Nat Photonics* **6**, 737-748 (2012).
2. Yang F, Cao W, Zheng GC, Qiu L, Nie ZH, Li Y. Plasmonic metasurfaces: Light-matter interactions, fabrication, applications and future outlooks. *Prog Mater Sci* **154**, 81 (2025).
3. Meinzer N, Barnes WL, Hooper IR. Plasmonic meta-atoms and metasurfaces. *Nat Photonics* **8**, 889-898 (2014).
4. Guo YJ, Xu ZY, Curto AG, Zeng YJ, Van Thourhout D. Plasmonic semiconductors: materials, tunability and applications. *Prog Mater Sci* **138**, 51 (2023).
5. Guan J, Park JE, Deng SK, Tan MJH, Hu JT, Odom TW. Light-Matter Interactions in Hybrid Material Metasurfaces. *Chem Rev*, 27 (2022).
6. Roth DJ, Krasavin AV, Zayats AV. Nanophotonics with Plasmonic Nanorod Metamaterials. *Laser Photon Rev* **18**, 17 (2024).
7. Nugroho FAA*, et al.* Inverse designed plasmonic metasurface with parts per billion optical hydrogen detection. *Nat Commun* **13**, 10 (2022).
8. Lee J*, et al.* Plasmonic biosensor enabled by resonant quantum tunnelling. *Nat Photonics* **19**, 938-945 (2025).
9. Wang YL*, et al.* Wearable plasmonic-metasurface sensor for noninvasive and universal molecular fingerprint detection on biointerfaces. *Sci Adv* **7**, 10 (2021).
10. John-Herpin A*, et al.* Metasurface-Enhanced Infrared Spectroscopy: An Abundance of Materials and Functionalities. *Adv Mater* **35**, 25 (2023).
11. Palermo G*, et al.* Plasmonic Metasurfaces Based on Pyramidal Nanoholes for High-Efficiency SERS Biosensing. *ACS Appl Mater Interfaces* **13**, 43715-43725 (2021).
12. Azzam SI*, et al.* Ten years of spasers and plasmonic nanolasers. *Light-Sci Appl* **9**, 21 (2020).
13. Gwo S*, et al.* Plasmonic Metasurfaces for Nonlinear Optics and Quantitative SERS. *ACS Photonics* **3**, 1371-1384 (2016).

14. Pavlov DV*, et al.* Nonlinear light conversion and infrared photodetection with laser-printed plasmonic metasurfaces supporting bound states in the continuum. *Light-Sci Appl* **15**, 11 (2026).
15. Tittl A*, et al.* Imaging-based molecular barcoding with pixelated dielectric metasurfaces. *Science* **360**, 1105-+ (2018).
16. Sain B, Meier C, Zentgraf T. Nonlinear optics in all-dielectric nanoantennas and metasurfaces: a review. *Adv Photonics* **1**, 14 (2019).
17. Koshelev K*, et al.* Subwavelength dielectric resonators for nonlinear nanophotonics. *Science* **367**, 288-+ (2020).
18. Khurgin JB. How to deal with the loss in plasmonics and metamaterials. *Nat Nanotechnol* **10**, 2-6 (2015).
19. Wang BQ*, et al.* High-Q Plasmonic Resonances: Fundamentals and Applications. *Adv Opt Mater* **9**, 30 (2021).
20. Liang Y*, et al.* Bound States in the Continuum in Anisotropic Plasmonic Metasurfaces. *Nano Lett* **20**, 6351-6356 (2020).
21. Sun SY*, et al.* Tunable plasmonic bound states in the continuum in the visible range. *Phys Rev B* **103**, 9 (2021).
22. Seo IC, Kim S, Woo BH, Chung I-S, Jun YC. Fourier-plane investigation of plasmonic bound states in the continuum and molecular emission coupling. **9**, 4565-4577 (2020).
23. Azzam SI, Shalaev VM, Boltasseva A, Kildishev AV. Formation of Bound States in the Continuum in Hybrid Plasmonic-Photonic Systems. *Phys Rev Lett* **121**, 6 (2018).
24. Hsu CW, Zhen B, Stone AD, Joannopoulos JD, Soljacic M. Bound states in the continuum. *Nat Rev Mater* **1**, 13 (2016).
25. Muhammad N, Su ZX, Jiang Q, Wang YT, Huang LL. Radiationless optical modes in metasurfaces: recent progress and applications. *Light-Sci Appl* **13**, 21 (2024).

26. Koshelev K, Lepeshov S, Liu MK, Bogdanov A, Kivshar Y. Asymmetric Metasurfaces with High-Q Resonances Governed by Bound States in the Continuum. *Phys Rev Lett* **121**, 6 (2018).
27. Luo M*, et al.* High-Sensitivity Optical Sensors Empowered by Quasi-Bound States in the Continuum in a Hybrid Metal-Dielectric Metasurface. *ACS Nano* **18**, 6477-6486 (2024).
28. Aigner A*, et al.* Plasmonic bound states in the continuum to tailor light-matter coupling. *Sci Adv* **8**, 8 (2022).
29. Liang Y, Tsai DP, Kivshar Y. From Local to Nonlocal High-Q Plasmonic Metasurfaces. *Phys Rev Lett* **133**, 7 (2024).
30. Yao J*, et al.* Nonlocal meta-lens with Huygens' bound states in the continuum. *Nat Commun* **15**, 8 (2024).
31. Shastri K, Monticone F. Nonlocal flat optics. *Nat Photonics* **17**, 36-47 (2023).
32. Chen Y, Fleury R, Seppecher P, Hu GK, Wegener M. Nonlocal metamaterials and metasurfaces. *Nat Rev Phys* **7**, 299-312 (2025).
33. Zakomirnyi VI, Rasskazov IL, Gerasimov VS, Ershov AE, Polyutov SP, Karpov SV. Refractory titanium nitride two-dimensional structures with extremely narrow surface lattice resonances at telecommunication wavelengths. *Appl Phys Lett* **111**, 4 (2017).
34. Kravets VG, Kabashin AV, Barnes WL, Grigorenko AN. Plasmonic Surface Lattice Resonances: A Review of Properties and Applications. *Chem Rev* **118**, 5912-5951 (2018).
35. Yu JB, Yao WZ, Qiu M, Li Q. Free-space high-Q nanophotonics. *Light-Sci Appl* **14**, 18 (2025).
36. Bin-Alam MS*, et al.* Ultra-high-Q resonances in plasmonic metasurfaces. *Nat Commun* **12**, 8 (2021).
37. Yang F*, et al.* Fabrication of Centimeter-Scale Plasmonic Nanoparticle Arrays with Ultranarrow Surface Lattice Resonances. *ACS Nano* **17**, 725-734 (2023).

38. Kelavuori J, Panahpour A, Huttunen MJ. Dispersion-induced Q-factor enhancement in waveguide-coupled surface lattice resonances. *Phys Rev B* **110**, 9 (2024).
39. Qi XY, Pérez LA, Alonso MI, Mihi A. High Q-Factor Plasmonic Surface Lattice Resonances in Colloidal Nanoparticle Arrays. *ACS Appl Mater Interfaces* **16**, 1259-1267 (2023).
40. Huang S*, et al.* High-Q multimodal guided-surface lattice resonances in index-discontinuous environments. *Nat Commun*, (2026).
41. Jia QW*, et al.* Reusable high-Q plasmonic metasurface. *Photonics Res* **13**, 1010-1020 (2025).
42. Allayarov I*, et al.* Strong coupling of collective optical resonances in dielectric metasurfaces. *Light-Sci Appl* **14**, 11 (2025).
43. Trinh QT*, et al.* Coexistence of surface lattice resonances and bound states in the continuum in a plasmonic lattice. *Opt Lett* **47**, 1510-1513 (2022).
44. Shen Y*, et al.* Ultrasmooth Gold Nanogroove Arrays: Ultranarrow Plasmon Resonators with Linewidth down to 2 nm and Their Applications in Refractive Index Sensing. *Adv Funct Mater* **32**, 12 (2022).
45. Tse JTY, Murai S, Tanaka K. Resonant Critical Coupling of Surface Lattice Resonances with a Fluorescent Absorptive Thin Film. *J Phys Chem C*, 10 (2023).
46. Sauvan C, Hugonin JP, Maksymov IS, Lalanne P. Theory of the Spontaneous Optical Emission of Nanosize Photonic and Plasmon Resonators. *Phys Rev Lett* **110**, 5 (2013).
47. Cao F*, et al.* Interaction of plasmonic bound states in the continuum. **11**, 724-731 (2023).
48. Fan SH, Suh W, Joannopoulos JD. Temporal coupled-mode theory for the Fano resonance in optical resonators. *J Opt Soc Am A-Opt Image Sci Vis* **20**, 569-572 (2003).
49. Piper JR, Liu V, Fan SH. Total absorption by degenerate critical coupling. *Appl Phys Lett* **104**, (2014).

50. Li HY*, et al.* Scalable Manufacturing of Low-Symmetry Plasmonic Nanospindle Arrays with Tunable Surface Lattice Resonance. *ACS Nano* **19**, 7391-7400 (2025).
51. Kelavuori J*, et al.* Thermal Control of Plasmonic Surface Lattice Resonances. *Nano Lett* **22**, 3879-3883 (2022).
52. Wang ZY, Ho YL, Cao T, Yatsui T, Delaunay JJ. High-Q and Tailorable Fano Resonances in a One-Dimensional Metal-Optical Tamm State Structure: From a Narrowband Perfect Absorber to a Narrowband Perfect Reflector. *Adv Funct Mater* **31**, 11 (2021).
53. Dao TD*, et al.* An On-Chip Quad-Wavelength Pyroelectric Sensor for Spectroscopic Infrared Sensing. *Adv Sci* **6**, 9 (2019).
54. Wang J, Weber T, Aigner A, Maier SA, Tittl A. Mirror-Coupled Plasmonic Bound States in the Continuum for Tunable Perfect Absorption. *Laser Photon Rev* **17**, 10 (2023).
55. Li GY, Du X, Xiong L, Yang XH. Plasmonic Metasurfaces with Quality Factors Up to 790 in the Visible Regime. *Adv Opt Mater* **11**, 9 (2023).

## Acknowledgements

The authors acknowledge the financial support from The National Key Research and Development Program of China (2023YFB3610800).

## Competing interests

The authors declare no competing interests.

# Supporting Information of Ultrahigh-$Q$ All-Metallic Metasurfaces with Robust Near-Perfect Absorption

Hanlei Xu[1], Han Qi[1], Chaoying Shi[1], Hong Zhang[1], Jun Wang[1], Ziming Meng[2], Guoliang Deng[1] and Hao Zhou[1]

[1]*College of Electronics and Information Engineering, Sichuan University, Chengdu 610065, China*

[2]*School of Physics and Optoelectronic Engineering, Guangdong University of Technology, Guangzhou 510006, China*

**Corresponding author(s). E-mail(s): Hao Zhou (zhoufirst@scu.edu.cn);*

*Guoliang Deng (gdeng@scu.edu.cn)*

**S1 The fabrication procedure of SDPs metasurface.**

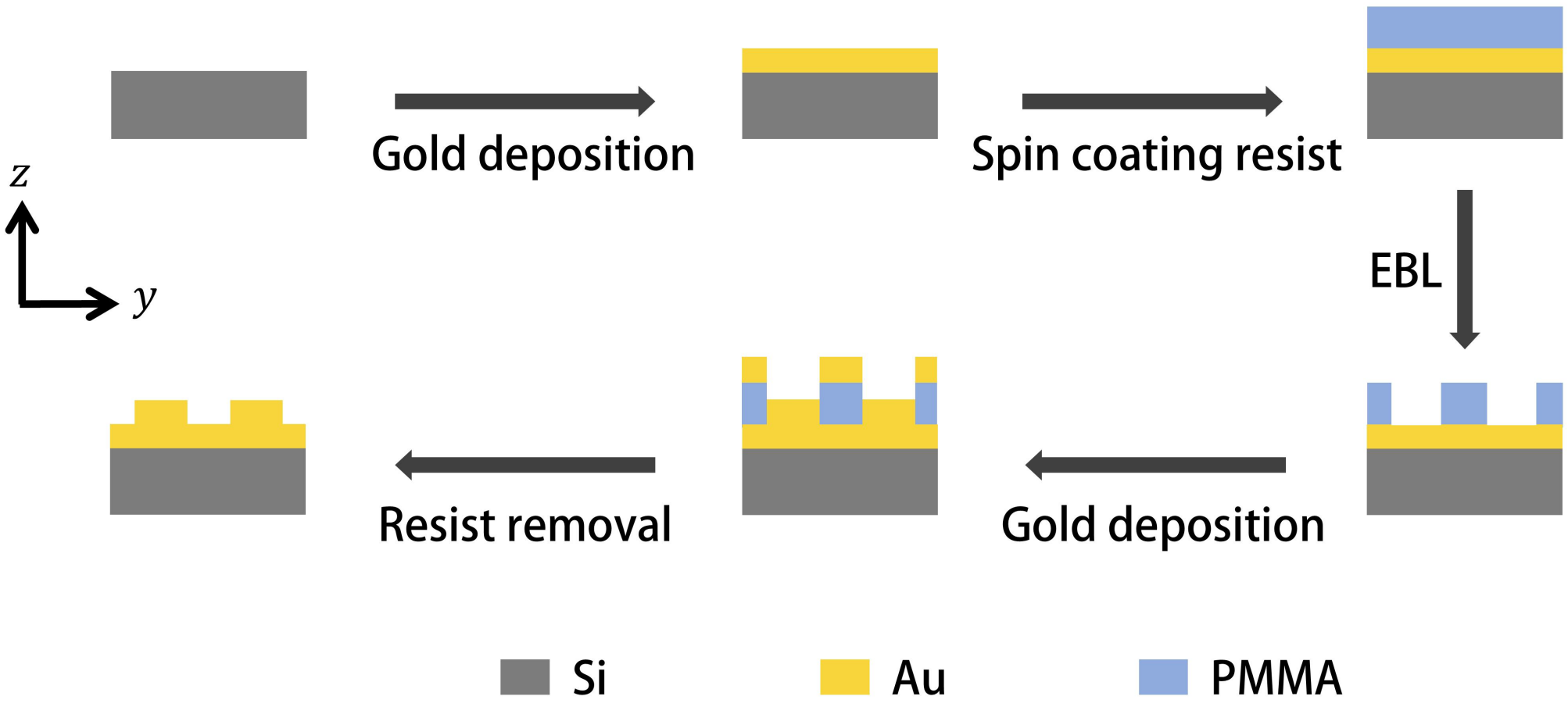

**Figure S1. Schematics showing the fabrication procedure of SDPs metasurface.**

The fabrication procedure is shown in Figure S1. The samples were fabricated using standard nanofabrication techniques, primarily involving electron-beam lithography (EBL), physical vapor deposition (PVD), and lift-off processes. First, an 80-nm-thick gold film was deposited onto a silicon substrate via thermal evaporation. Subsequently, a layer of EBL photoresist was spin-coated onto the gold film, followed by patterning using EBL. Next, a second 80-nm-thick gold layer was deposited on the developed sample using thermal evaporation. Finally, a lift-off process was performed to remove the remaining resist and obtain the final samples.

## S2 experimental setup for measuring ultrahigh $Q$ factors.

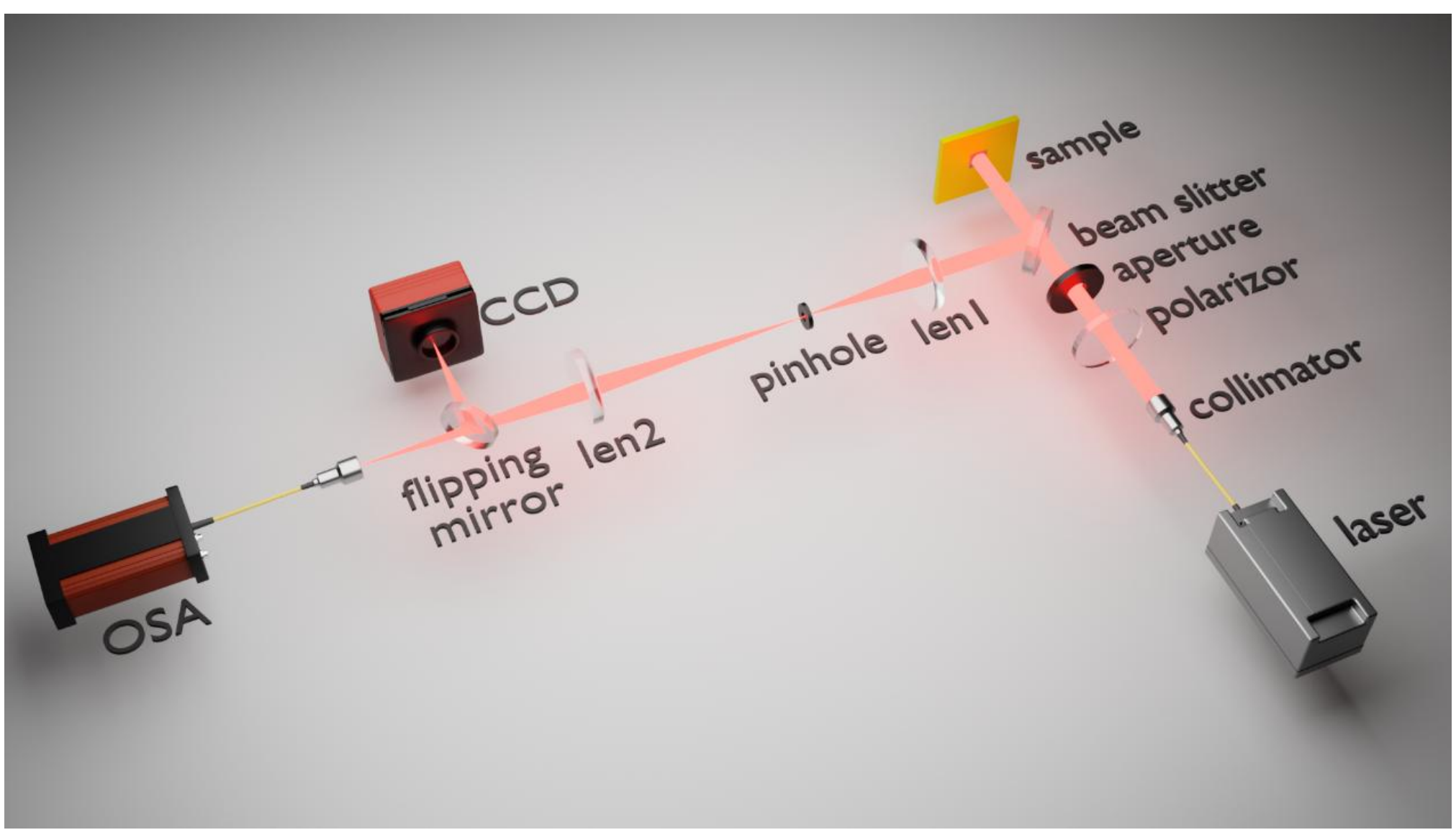

**Figure S2. Schematic of the home-built optical setup for high-resolution reflection spectroscopy.** A collimated laser beam passes through a polarizer and beam-shaping optics before being focused onto the fabricated plasmonic metasurface. The reflected signal is collected through the same optical path, redirected by a flipping mirror, and recorded by a CCD or optical spectrum analyzer (OSA), enabling high-resolution characterization of ultranarrow resonances and high-$Q$ factors.

## S3 Bright mode analysis

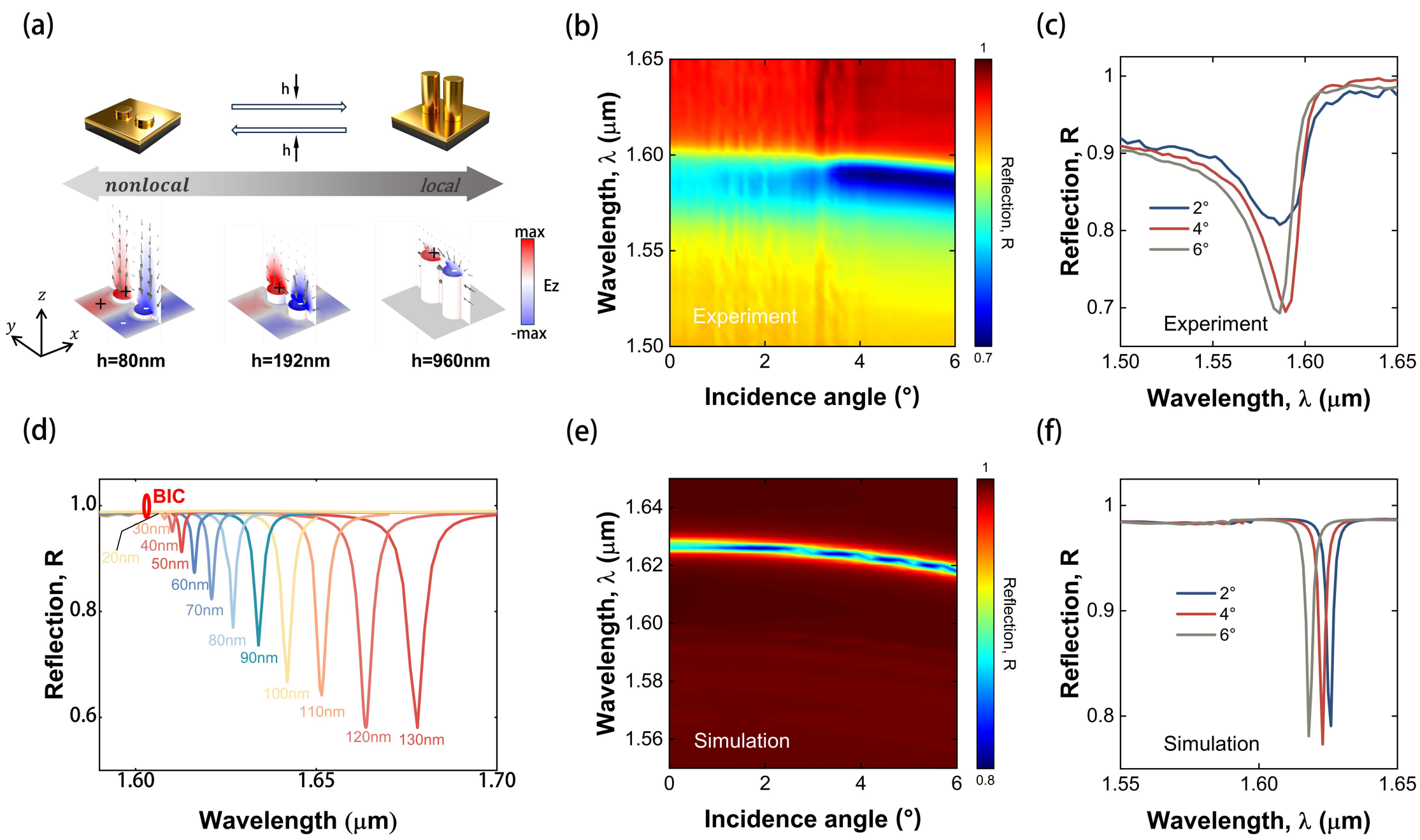

**Figure S3 Local-to-nonlocal evolution and angle-resolved optical response of the bright mode. (a)** Evolution of the electric field $E_z$ during the transition of the bright mode from the local to the nonlocal regime at different heights ($h = 80$ nm, $h = 192$ nm, and $h = 960$ nm). **(b)** Experimental and **(e)** Simulated angle-resolved reflection spectrum for the bright mode under TE polarization. **(c)** Experimental and **(f)** Simulated reflection spectra under TE polarization for different incident angles (2°, 4°, and 6°). **(d)** Evolution of the reflection spectra as the height $h$ changes. When $h$ is small, the bright mode approaches the dark mode.

The bright mode corresponds to an odd-symmetry in-plane plasmonic resonance and can be directly excited under TE-polarized incidence. Fig. S3(a) shows the evolution of its $E_z$ field distribution as the pillar height decreases from the local regime ($h = 960$ nm), through an intermediate regime ($h = 192$ nm), to the nonlocal regime ($h = 80$ nm). For large $h$, the bright mode is strongly confined near the pillar tops, showing pronounced localized hotspots. As $h$ decreases, the field becomes progressively delocalized and extends over neighboring unit cells, indicating the emergence of a nonlocal collective response. Figs S3(b, c) and S3(e, f) show the experimental and simulated angle-resolved reflection spectra of the bright mode under TE-polarized incidence, together with the spectra at selected incident angles of 2°, 4°, and 6°. The good agreement between experiment and simulation confirms the reliability of the numerical model and verifies the high-$Q$ bright-mode response in the nonlocal regime. Fig. S3(d) further presents the normal-incidence reflection spectra for different pillar heights, showing the continuous spectral evolution of the bright mode as $h$ changes.

Although the bright mode also exhibits a local-to-nonlocal evolution, it is not the relevant branch for realizing the ultrahigh-$Q$, near-critical-coupling hybrid resonance in the main text. Fig. S3(e) shows that, under TE-polarized excitation, the bright-mode resonance remains weakly absorbing and does not develop a pronounced RA-mediated hybrid feature in the present design. In contrast, the dark FP-BIC mode provides a radiation-suppressed state with a much larger $Q_{\mathrm{rad}}$, making it more suitable for deliberately introducing and tuning radiative leakage through the RA-associated channel. Therefore, the bright mode is discussed here as a comparative reference, whereas the hybrid resonance in the main text arises from the coupling between the dark FP-BIC mode and the RA-associated lattice resonance.

## S4. Refractive-index sensing in aqueous glucose solutions

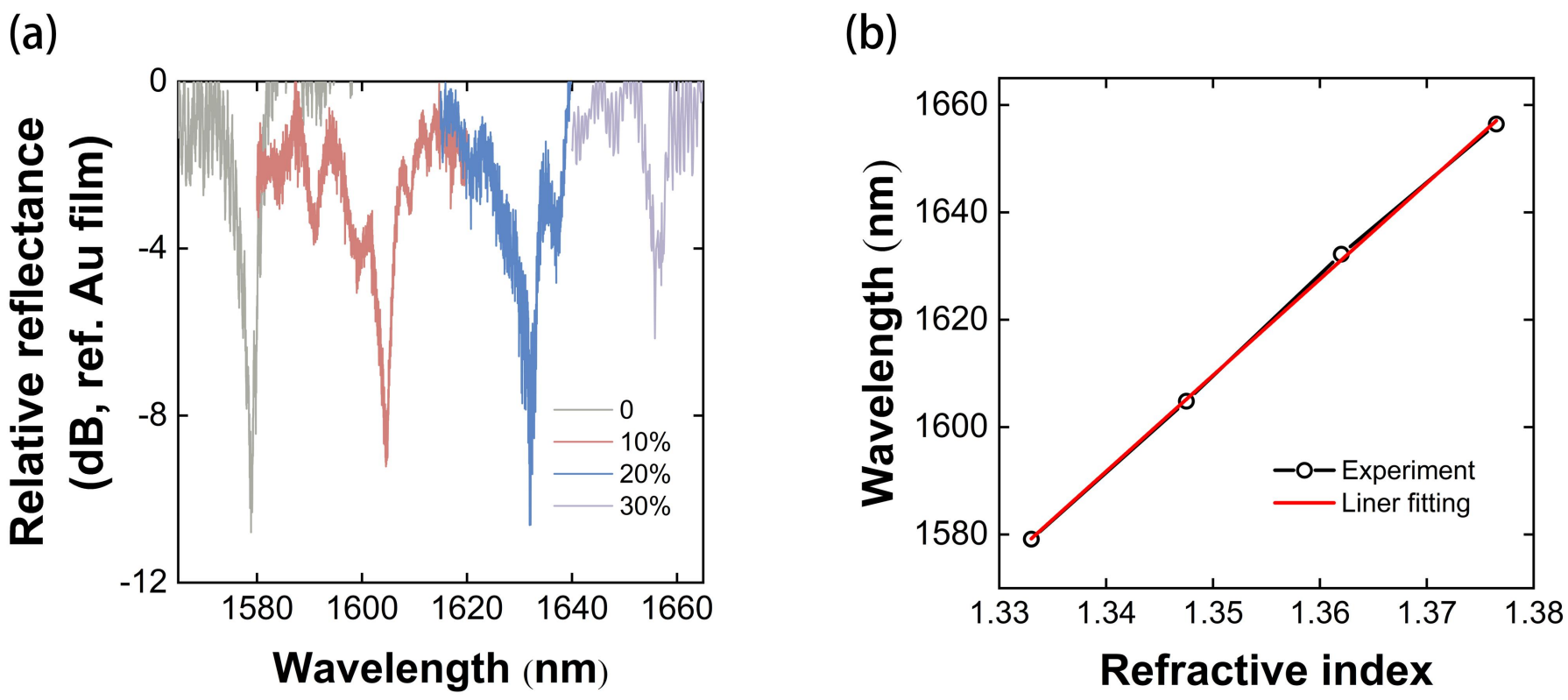

**Figure S4. Glucose-concentration sensing response of the SDPs metasurface in aqueous solutions. (a)** The experimental reflection spectra of the SDPs metasurface in a liquid environment when the RI increasing from n = 1.333 to n = 1.3833 under normal TM polarized incidence. **(b)** Dependence of the shift of resonance wavelength on the RI of environmental medium in a liquid environment.

To demonstrate the potential application of the proposed SDPs metasurface, we experimentally verified its refractive-index (RI) sensing performance in a liquid environment. The bulk RI sensitivity, defined as the resonance shift induced by variations in the RI of the environmental medium, is expressed as $S = \Delta\lambda_{res}/\Delta n_{env}$, where $\Delta\lambda_{res}$ and $\Delta n_{env}$ denote the shift in resonant wavelength and the change in environmental refractive index, respectively. To ensure the resonant wavelength falls within the detectable range of our existing laboratory setup (around 1600 nm), the lattice period of the metasurface was set to $P = 1200\ nm$, while the other structural parameters were proportionally scaled except for the pillar height. Glucose aqueous solutions with mass fractions of 0%, 10%, 20%, and 30% were used as analytes, corresponding to refractive indices of 1.333, 1.3455, 1.3620, and 1.3833, respectively.

Fig. S4(a) shows the measured reflection spectra under normal TM-polarized incidence as the environmental RI increases from 1.333 to 1.3833. The resonance exhibits a clear redshift with increasing RI, as summarized in Fig. S4(b). The nearly linear dependence of the resonance wavelength on the environmental RI enables quantitative analysis of the analyte concentration. From the linear fitting, the bulk RI sensitivity of the SDP metasurface in the liquid environment is extracted to be $S = 1700nm/RIU$. These results demonstrate the potential of the proposed high-$Q$ plasmonic metasurface for label-free RI sensing and glucose-concentration detection based on refractive-index modulation.

## S5. Comparison of this work with representative plasmonic metasurface works.

| Ref. (main text) | Mechanism | Q-factor | ΔI | λ(nm) | Structure | $\theta_{incident}$ | Superstrate |
|---|---|---|---|---|---|---|---|
| **36** | SLR | 2340 | 60% | 1550 | Au nanoparticles | Normal | Silica |
| **41** | Hybrid plasmonic-photonic SP-BIC | 603 | 50% | 1585 | Au nanoparticles | 0.9° | Silica |
| **37** | SLR | 218 | 50% | 870 | Au nanoparticles | Normal | Si |
| **39** | SLR | 270 | 70% | 1150 | Au nanoparticles | Normal | glass |
| **50** | SLR | 309 | 30% | 1205 | Au nanoparticles | Normal | Si |
| **44** | FP-WA coupling | 350 | 51% | 702 | Au nanogrooves | Normal | / |
| **28** | q-BIC | 180 | 80% | 8300 | Au nanofins | 10° | / |
| **20** | q-BIC | 50 | 80% | 4977 | Au nanoparticles | 9° | / |
| **51** | SLR | 750 | 40% | 1100 | Al nanoparticles | Normal | glass |
| **52** | Tamm plasmon polariton assisted Fano resonance | 1120 | 92% | 1150 | Au-DBR multilayer | Normal | Si |
| **53** | SLR | 73 | 99% | 3722 | Au nanoparticles | Normal | Si3N4 |
| **54** | q-BIC | 45 | 80% | 6600 | Au nanobars | Normal | air |
| **55** | SLR | 790 | 46% | 712 | Au nanoparticles | 25° | Silica |
| **this work** | BIC-RA coupling | 2180 | 99% | 1600 | Au symmetric double-pillars | Normal | / |